\documentclass{iaus}
\usepackage{graphicx}

\def\Ha{H$\alpha$}
\def\kms{{\rm\,km\,s^{-1}}}
\def\hii{{\sc H\thinspace ii}}
\def\hi{{\sc H\thinspace i}}

\def\spose#1{\hbox to 0pt{#1\hss}}
\def\Dt{\spose{\raise 1.5ex\hbox{\hskip3pt$\mathchar"201$}}}    
\def\mdot{\Dt{M}}

\title[Superbubble Evolution] 
{Towards Resolving the Evolution of Multi-Supernova Superbubbles}

\author[M. S. Oey]   
{M. S. Oey$^1$}

\affiliation{$^1$Department of Astronomy, University of Michigan, 830
Dennison Building, Ann Arbor, MI\ \ \ 48109-1042, USA}

\pubyear{2006}
\volume{237}  
\pagerange{119--126}
\date{?? and in revised form ??}
\jname{Triggered Star Formation in a Turbulent ISM}
\editors{J. Palou\v s \& B.G. Elmegreen, eds.}
\begin{document}

\maketitle

\begin{abstract}
Interstellar superbubbles generated by multiple supernova explosions
are common in star-forming galaxies.  They are the most obvious
manifestation of mechanical feedback, and are largely responsible for
transferring both thermal and kinetic energy to the interstellar
medium from the massive star population.  However, the details of this
energy transfer remain surprisingly murky when individual objects are
studied.   I will summarize what we currently know about candidate
dominant processes on these scales.  
\end{abstract}

\firstsection 
\section{Introduction}

The previous contributions showed a wealth of
\hi\ shells and supershells.  How well-established is it, that these
structures originate from mechanical feedback, namely, winds and
supernovae (SNe) from OB associations?  If we are to understand
how massive star feedback affects galaxies on global scales, then it
is essential that we understand the origin and evolution of
superbubbles, which are the direct manifestation of feedback.  We
refer not only to mechanical energy, but also to
photoionization, and dispersal of nucleosynthesis products from the SNe
and massive stars.  In this presentation, I use the definition
introduced by You-Hua Chu in the 1980's:  a ``superbubble'' is
a shell structure originating from multiple stars, rather than a
definition related to its size.   

The default model for superbubbles is the standard, adiabatic
evolution that assumes a hot ($10^6+$ K), shock-heated interior whose
pressure drives the outer shell growth (Pikel'ner 1968; Castor et
al. 1975; Dyson 1977).  For this model, the shell parameters are
determined by only three input parameters:  the mechanical
``luminosity'' $L$, the ambient density $n$, and the age $t$, e.g., 
\begin{equation}\label{eq_R}
R\propto (L/n)^{2/5}\ t^{3/5} \quad ,
\end{equation}
\begin{equation}\label{eq_v}
v=\frac{dR}{dt}\propto (L/n)^{2/5}\ t^{-2/5} \quad .
\end{equation}
We also must know conditions in the ambient environment, for
example, the ambient pressure and density distribution.  The above is
the simplest, analytic representation of this model, but there is also
a large body of work in hydrodynamic simulations of these objects over
the past two decades (e.g., Mac Low et al. 1989; Bisnovatyi-Kogan et
al. 1989; Palou\v s et al. 1990; Tenorio-Tagle et al. 1991;
Tomisaka 1990; Silich et al. 1996; Gazol-Pati\~no \& Passot 1999;
Strickland \& Stevens 2000).

What evidence do we seek, that ``star formation,'' i.e., massive stars,
are responsible for the ubiquitous superbubbles that we see in
gas-rich galaxies?  Since the adiabatic model is based on the
shock-heated interior, we expect gas at multiple temperature phases,
including hot, X-ray emitting gas; warm, $10^4$ K photoionized gas; and
also the neutral gas that we have seen.  Second, we should also
find that objects whose input parameters are known
should follow the prescribed evolution above.  Third, the
statistical properties of the superbubble populations
should be consistent with the
statistical properties of the putative parent star-forming regions.
Fourth, and perhaps most obvious, we expect a one-to-one
correspondence between the superbubbles and the massive star
clusters. 

\section{Multi-phase ISM}

As mentioned by Elias Brinks in his talk, we often do
see the existence of gas at the different temperatures that are
predicted by the adiabatic model.  A recent example is the
X-ray emission in N51 D (Cooper et al. 2004), along with photoionized
\Ha\ emission.  Superbubbles were already detected in X-rays
by {\sl Einstein} and {\sl ROSAT} (e.g., Chu \&
Mac Low 1990; Wang \& Helfand 1991).  It has long been known that
there are two categories of X-ray emission from these objects:  X-ray
bright, and X-ray dim (Chu et al. 1995).  The former show X-ray
emission in excess of what is expected from the adiabatic model, and 
the enhancements are likely caused by secondary SN
blastwave impacts to the shell walls (Chu \& Mac Low 1990; Oey 1996).

H recombination emission is generally consistent with photoionization
by the observed early-type stars, although density-bounding and
shock-heating contributions are also often factors (Hunter et
al. 1995; Oey et al. 2000).  We usually see \Ha\ emission on the
interior of \hi\ shells, consistent with the model that 
star formation is responsible for the superbubbles.  Kim et
al. (1999) studied the \hi\ shells of the Large Magellanic
Cloud (LMC) and were even able to suggest an evolutionary sequence
defined by relative \Ha\ and \hi\ morphology.

Intermediate-temperature ions are also present.  {\sc C iv} and Si
{\sc iv}, are often seen in absorption in the lines of sight toward
massive stars within superbubbles (Chu et al. 1994).  Recently, {\sc O
vi} was also detected by {\sl FUSE} in the line of sight
toward one LMC superbubble (Danforth \& Blair 2006). 

\section{Dynamics of Individual Superbubbles}

If the input mechanical power, ambient conditions, and age of the
superbubbles can be determined, we can test for consistency
with the adiabatic evolution.  Such studies are possible for objects in the
Milky Way and Magellanic Clouds, where the stellar population can be
resolved, and these studies invariably show the existence of a growth-rate
discrepancy such that the shells are much smaller than predicted by
the apparent input parameters (Saken  et al. 1992; Brown  et al. 1995;
Oey \& Massey 1995; Hunter et al. 1995; Oey 1996; Cooper et al. 2004).
The problem has been known for many
years, and is even seen in single-star nebulae generated by Wolf-Rayet
stars (e.g., Cappa et al. 2001, 2005).

What is the current status in resolving this growth-rate discrepancy?
There are a number of possible important factors.  Those discussed
below are the most likely candidates.  Combinations of multiple
effects may also be at play, including others, such as
viscous drag (I. Goldman, private communication), that we do not
discuss fully below.

\subsection{Input power overestimated?}

The first possibility for resolving the superbubble growth-rate
discrepancy is that the input mechanical luminosity has been
overestimated.  Stellar wind mass-loss rates $\mdot$ are
especially suspect, since it has long been suggested that clumping in
the winds leads to overestimates in $\mdot$ from radio continuum
measurements (e.g., Hillier 1991; Nugis et al. 1998).
In recent years, this appears to be confirmed by X-ray 
line profile fitting (Cohen et al. 2006; Miller et al. 2002;
Kramer et al. 2003).  Fullerton et al. (2006) also find the same
result from {\sc P v} line profiles observed by {\it FUSE}.  Their
analysis of this dominant ion suggests that the overestimates in
$\mdot$ may be as high as two orders of magnitude in some cases!
These overestimates in $\mdot$ are
relevant primarily to the youngest superbubbles, whose evolution is
still dominated by stellar winds instead of SNe, which applies to most
dynamical studies of optical or X-ray selected superbubbles.

\subsection{Ambient density underestimated?}

As seen in equations~\ref{eq_R} and \ref{eq_v}, an underestimate in
the ambient density has an equivalent effect to an overestimate in
$L$.  Oey et al. (2002) mapped the immediate environment of three
\Ha-selected LMC superbubbles to determine whether the neutral gas
environment was unusually dense.  We found an extreme range in
conditions for the three objects, with one essentially in an
\hi\ void, another nestled amongst a number of \hi\ clouds, and the
third in a region with no obvious relation to the observed \hi.
This lack of any systematic effect suggests that the ambient density
may not be the primary source of the growth-rate discrepancy, although
it also demonstrates that the ambient environment is more complex
than assumed.  

We also note that if a superbubble originates in a higher-density
cloud, then a mini-blowout from the cloud can also accelerate the
shell's observed expansion velocity relative to its radius
(Oey \& Smedley 1998; Mac Low et al. 1998).
Hence it is possible
to reproduce unusual observed kinematics in particular objects.

\subsection{Ambient pressure underestimated?}

Another important environment parameter is the interstellar pressure,
which counteracts the growth of the shells.  In general, models assume
a fiducial ambient $P/k$ on the order of $10^4\ \rm cm^{-3}\ K$ or
less.  The various sources of ambient pressure can be broadly
described as the thermal pressure, magnetic pressure, turbulent
pressure, and cosmic ray pressure.  We note that all of these
correlate with star-formation rate (SFR).  Likewise, superbubbles tend
to be prominent in active star-forming galaxies like the LMC and
localized active environments.  Indeed, the very existence of
star-formation in these regions implies higher local pressure (M. Mac Low,
private communication).  It is therefore plausible that the
ambient pressure for superbubbles may have been systematically
overestimated. 

Oey \& Garc\'\i a-Segura (2004) discuss this possibility and present 2-D
hydrodynamic simulations of six LMC superbubbles for ambient $P/k =
10^4$ and $10^5\ \rm cm^{-3}\ K$.  The models are generated with
the same input mechanical luminosity that is estimated from the observed
stellar population by Oey (1996).  For the lower pressure, the simulated
radial density profiles show an extended photoionized morphology that
is inconsistent with the observed \Ha\ data.
In contrast, the
simulations assuming ambient $P/k = 10^5\ \rm cm^{-3}\ K$ show no
extended \Ha\ emission, and an \Ha\ morphology that agrees well
with the observations.
The observed and modeled velocity structures are similarly more
consistent with the models for the high-pressure environment in all
cases.  The \hi\ radial profile predictions also differ, and can be
used to further test these models.

\subsection{Radiative Cooling?}

If the hot superbubble interiors are in fact losing
energy by radiative cooling, then the growth will not keep pace with
the adiabatic model.  
Mass-loading into the hot interior via evaporation from the shell
walls or ablation of clouds and clumps can drive radiative cooling
(e.g., Hartquist et al. 1986; Arthur \& Henney 1996).
Alternatively, an increase in metallicity due to the injection of
nucleosynthesis products from the parent SNe can also enhance the
cooling rate.  Silich \& Oey (2002) show that the X-ray luminosity of
a low-metallicity ($Z=0.05Z_\odot$) superbubble can be increased by
almost an order of magnitude, simply by products from 3 -- 4 SNe.
Silich et al. (2001) examine this issue for starburst superwinds.
However, the observed X-ray luminosities generally do not appear to
suggest anomalous cooling from the superbubble interiors (e.g., Chu et
al. 1995).

\subsection{Energy Transferred to Cosmic Rays?}

Another factor whose importance has been
underemphasized, is the transfer of superbubble energy to 
cosmic rays.  Multi-SN superbubbles are especially
efficient at accelerating cosmic rays because the blastwaves expand into a
pre-heated environment (e.g., Parizot et al. 2004).  The
superbubble interiors thus harbor strong MHD turbulence and
magnetic fields (e.g., Bykov 2001; Bykov \& Toptygin 1987), which are
needed for cosmic ray acceleration.  Because of the multiple SNR shocks,
superbubbles also promote multiple accelerations, which can 
push cosmic rays to higher energies (Parizot et al. 2004; Klepach
et al. 2000).  The cosmic ray energy distributions and isotope
abundances are broadly consistent with superbubble origins.  While the
role of superbubbles in explaining cosmic ray properties has been
recognized for some time, the effect of energy transfer on the parent
objects themselves has only been discussed recently, and rough
estimates suggest that a few tenths of superbubble kinetic energy
could be lost to cosmic rays (Parizot et al. 2004; Bykov
2001).  For individual young SNRs, simulations predict growth
deviations $\gtrsim 10$\% and reduced X-ray luminosities
(Ellison et al. 2004).  This energy sink applies more to
SN-dominated superbubbles, rather than stellar wind-dominated
objects. 

\section{Properties of global mechanical feedback}

Another approach to evaluating the adiabatic model is
to compare observations and predictions for the statistical properties
of superbubbles in galaxies, based on the known global star-formation
properties.  Oey \& Clarke (1997) derived size distributions and
expansion velocity distributions for extremes in
star-formation history and star-cluster mass functions.  The latter,
which produces the \hii\ region luminosity function, is
generally a robust power law with a dependence of $L^{-2}$ in
differential form (e.g., Efremov \& Elmegreen 1997; Oey \& Clarke
1998).  This yields a differential size distribution,
\begin{equation}\label{eq_NR}
N(R)\ dR\propto R^{-3}\ dR 
\end{equation}
and corresponding distribution in expansion velocities,
\begin{equation}
N(v)\ dv\propto v^{-7/2}\ dv \quad .
\end{equation}
We find excellent agreement with these relations for the \hi\ shell
population found by Staveley-Smith et al. (1997) for the SMC.
More recently, Hatzidimitriou et al. (2005)
updated the SMC shell catalog and re-examined these relations, inferring
a different star-formation history, but one that is still
consistent with a star-formation origin for the structures, plus
adiabatic evolution.  \hi\ shell size distributions for a few other
galaxies also have been examined, for example, the Milky Way
(S. Ehlerov\`a, these proceedings), LMC (Kim et al. 1999), M31, M33,
Holmberg~II (Oey \& Clarke 1997), and NGC 2403 (Thilker et al. 1998;
Mashchenko et al. 1999).  These other catalogs are generally
consistent with equation~\ref{eq_NR}, but the statistics are
more incomplete.

Analysis of the global size distributions leads to the definition
of a critical star-formation rate, above which the shells merge
and shred the neutral ISM, generating
pressure-driven outflows from their parent galaxy disks (Clarke \& Oey 2002):
\begin{equation}\label{eq_sfrcrit}
\rm SFR_{crit} = 0.15\ M_{ISM,10}\ \sigma_{{\it v},10}^2 / {\it f_d} \quad
	M_\odot\  yr^{-1} \quad ,
\end{equation}
where M$_{\rm ISM,10}$ and $\sigma_{v,10}$ are the mass of the ISM in
units of $10^{10}\ \rm M_\odot$ and the thermal velocity dispersion in
units of $10\ \kms$, respectively.  The geometric correction factor
$f_d$ is on the order of unity.  Is this expected shredding of the
neutral ISM for high SFR morphologically apparent in the \hi\
datasets?  

We can compare the 
\hi\ maps and shell catalogs for the LMC (Kim et al. 1999) and SMC
(Staveley-Smith et al. 1997).  The two surveys were both
carried out with the Australia Telescope, at similar
depth and spatial resolution.  Although the LMC is a larger galaxy and
has a much higher star-formation rate than the SMC, the number of
coherent \hi\ shells identified in the LMC survey is only 126,
roughly one-quarter of the 509 (Hatzidimitriou et al. 2005)
found in the SMC.  Ordinarily, we would expect a much {\it larger}
number of shells in the LMC than in the
SMC.  Morphologically, there is a noticeable contrast between these two
galaxies:  the LMC \hi\ has a filamentary appearance, consistent with
shredding and compression, whereas the SMC has a more quiescent, smoother
appearance.  It seems surprising that the SMC dataset yields a so much
larger shell catalog than the LMC.
It turns
out that for the LMC,\  $\rm SFR/SFR_{crit}\sim 1$,
whereas for the SMC,\  $\rm SFR/SFR_{crit}\sim 0.1$
(Oey 2001).  It will be interesting to examine shell catalogs and
properties for larger samples of galaxies, from
the THINGS \hi\ survey (E. Brinks, these proceedings), for example.

\section{Detailed correspondence with star-forming regions}

Last, but not least, we seek a one-to-one correspondence between the
observed superbubbles and parent star clusters.  To date, 
results are not as clear-cut as is desireable, but there
does appear to be broad consistency, though somewhat controversial.
Kim et al. (1999) find enough correspondence between the LMC \hi\ and \Ha\
data to suggest an evolutionary sequence based on the morphological
and kinematic relationship between these.  Hatzidimitriou et
al. (2005) examine the spatial relationship between the \hi\ shells
in the SMC and existing data for young stellar clusters,
and do find broad correspondence, although the quantitative
significance of the correlations is ambiguous.  They also find that
about 10\% of the \hi\ shells show no counterparts in the stellar
population.  In Holmberg~II, a targeted search for clusters in
the \hi\ shells failed to find the expected correspondence at optical
wavelengths (Rhode et al. 1999), although \Ha\ and FUV observations
seem more consistent with a mechanical feedback origin
(Stewart et al. 2000).

Finally, in the theme of this Symposium,
triggered star-formation is also an association of
massive star clusters with the creation of superbubbles.  There are
many examples of two-stage, sequential star formation; some 
well-known examples are N11 (Walborn \& Parker 1992), N44
(Oey \& Massey 1995), and N51 D (Oey \& Smedley 1998; Cooper et
al. 2004) in the LMC, and the Rosette Nebula
(Williams et al. 1995) in the Galaxy, among
others.  While it is difficult to establish a {\it causal}
relationship between two-stage star-forming regions, a three-stage
sequence is much more convincing.  We recently identified the Perseus
W3/4 complex in the Galaxy, identified as a triggered system by 
Thronson et al. (1985), as three-stage, hierarchical star formation, 
whose morphology is difficult to interpret as anything other than
causal, triggered star formation (Oey et al. 2005).  

\section{Summary}

To summarize, we see that observations are broadly consistent with
mechanical feedback from the most massive stars being responsible for
the formation of most superbubbles and shell structures seen in
star-forming galaxies, and that these objects can be understood in
terms of the standard, adiabatic evolution driven by massive star
winds and supernovae.  There is empirical, multi-wavelength
confirmation of the different temperature phases associated with the
shock heating of superbubble interiors and also photoionization by OB
stars in the youngest objects, as is qualitatively predicted by the
adiabatic model.  Statistical properties of shell populations are
quantitatively consistent with predicted distributions in size and
expansion velocity, based on known global properties of the parent
star-forming regions and young star clusters.  Above a threshold
star-formation rate, we predict interaction among the shells and
shredding of the ISM, which appears to be observed in the contrasting
\hi\ morphology and shell populations between the LMC and SMC.  The expected
one-to-one correlation between young massive-star clusters and
superbubbles remains somewhat ambiguous, although a general
consistency is tentatively seen.

When comparing the observed evolution and kinematics of individual
objects with standard predictions based on detailed knowledge of the
parent stellar population and other input parameters, we find that the
objects are also broadly consistent with adiabatic evolution, but that
the shells are invariably smaller than expected.  There are a number
of possible reasons for this growth-rate discrepancy.  (1) The
assumed input power may be overestimated, especially with respect to
measured mass-loss rates for stellar winds.  (2) It is possible that
the ambient density has been systematically underestimated, although
our resolved observations of \hi\ environments for three objects does
not especially support this interpretation.  (3) There may be a
systematic underestimate of the ambient pressure, since the various
contributors to interstellar pressure all correlate with
star-formation activity.  (4) Enhanced radiative cooling may be
occurring in the hot superbubble interiors, caused by
mass-loading or metallicity enhancements from the parent SNe.
(5)  Finally, a somewhat-overlooked mechanism is the
transfer of mechanical energy from the superbubbles to
cosmic rays, which is plausible since superbubbles are an especially
effective acceleration environment.

\begin{acknowledgments}
Many thanks to the Symposium organizers and participants, with whom I
enjoyed many discussions.  This work was supported in part by NSF
grant AST-0448893. 
\end{acknowledgments}


\begin{thebibliography}{}
\bibitem[]{} Arthur, S. J. \& Henney, W. J. 1996, ApJ, 457, 752
\bibitem[]{} Bisnovatyi-Kogan, G. S., Blinnikov, S. I., \& Silich,
  S. A. 1989, Ap\&SS 154, 229
\bibitem[]{} Brown, A. G. A., Hartmann, D., \& Burton, W. B. 1995,
  A\&A, 300, 903
\bibitem[]{} Bykov, A. M. 2001, Space Sci. Rev. 99, 317
\bibitem[]{} Bykov, A. M. \& Toptygin, I. N. 2001, Astron. Lett. 27, 625
\bibitem[]{} Cappa, C., Niemela, V. S., Mart\'\i n, M. C., \&
  McClure-Griffiths, N. M. 2005, A\&A 436, 155
\bibitem[]{} Cappa, C. E., Rubio, M., \& Goss, W. M. 2001, AJ 121, 2664
\bibitem[]{} Castor, J., McCray, R., \& Weaver, R. 1975, ApJ 200, L107
\bibitem[]{} Chu, Y.-H., Chang, H.-W., Su, Y.-L., \& Mac Low, M.-M.,
	1995, ApJ, 450, 157
\bibitem[]{} Chu, Y.-H. \& Mac Low, M-M., 1990, ApJ, 365, 510
\bibitem[]{} Chu, Y.-H., Wakker, B., Mac Low, M.-M., \& Garc\'\i a-Segura, G.
	1994, AJ, 108, 1696
\bibitem[]{} Clarke, C. J. \& Oey, M. S. 2002, MNRAS, 337, 1299
\bibitem[]{} Cohen, D., Leutenegger, M. A., Grizzard, K. T., Reed,
  C. L., Kramer, R. H., \& Owocki, S. P. 2006, MNRAS 368, 1905
\bibitem[]{} Cooper, R. L., Guerrero, M. A., Chu, Y.-H., Chen,
  C.-H. R., \& Dunne, B. C. 2004, ApJ 605, 751
\bibitem[]{} Danforth, C. W. \& Blair, W. P. 2006, ApJ 646, 205
\bibitem[]{} Dyson, J. E. 1977, A\&A 59, 161
\bibitem[]{} Efremov, Y. N. \&  Elmegreen, B. G. 1997, ApJ 480, 235
\bibitem[]{} Ellison, D. C., Decourchelle, A., \& Ballet, J. 2004,
  A\&A 413, 189
\bibitem[]{} Fullerton, A. W., Massa, D. L., \& Prinja, R. K. 2006,
  ApJ 637, 1025
\bibitem[]{} Gazol-Pati\~no, A. \& Passot, T. 1999, ApJ 518, 748
\bibitem[]{} Hatzidimitriou, D., Stanimirovi\'c, S., Maragoudaki, F.,
  Staveley-Smith, L., Dapergolas, A., \& Bratsolis, E. 2005, MNRAS
  360, 1171
\bibitem[]{} Hillier, D. J. 1991, A\&A 247, 455
\bibitem[]{} Hunter, D. A., Boyd, D. M., \& Hawley, W. N. 1995, ApJS
  99, 551
\bibitem[]{} Kim, S., Dopita, M. A., Staveley-Smith, L., \& Bessell,
	M. S. 1999, AJ, 118, 2797
\bibitem[]{} Klepach, E. G., Ptuskin, V. S., \& Zirakashvili,
  V. N. 2000, Astroparticle Phys., 13, 161
\bibitem[]{} Kramer, R. H., Cohen, D. H., \& Owocki, S. P. 2003, ApJ
  592, 532
\bibitem[]{} Mac Low, M.-M., Chang, T. H., Chu, Y.-H., Points, S. D., Smith,
	R. C., \& Wakker, B. P. 1998, ApJ, 493, 260
\bibitem[]{} Mac Low, M.-M., McCray, R., \& Norman, M. L. 1989, ApJ, 337, 141
\bibitem[]{} Mashchenko, S. Y., Thilker, D. A., \& Braun, R. 1999, A\&A, 343,
	352
\bibitem[]{} Miller, N. A., Cassinelli, J. P., Waldron, W. L.,
  MacFarlane, J. J., \& Cohen, D. H. 2002, ApJ 577, 951
\bibitem[]{} Nugis, T., Crowther, P. A., \& Willis, A. J. 1998, A\&A
  333, 956
\bibitem[]{} Oey, M. S. 1996, ApJ, 467, 666
\bibitem[]{} Oey, M. S. \& Clarke, C. J. 1997, MNRAS, 289, 570
\bibitem[]{} Oey, M. S. \& Clarke, C. J. 1998, AJ, 115, 1543
\bibitem[]{} Oey, M. S., Clarke, C. J., \& Massey, P. 2001, in {\sl
	Dwarf Galaxies and Their Environment,} eds. K. S. de Boer,
	R.-J. Dettmar, \& U. Klein, Shaker Verlag, 181.
\bibitem[]{} Oey, M. S., Dopita, M. A., Shields, J. C., \& Smith,
  R. C. 2000, ApJS, 128, 511
\bibitem[]{} Oey, M. S. \& Garc\'\i a-Segura, G. 2004, ApJ 613, 302
\bibitem[]{} Oey, M. S., Groves, B., Staveley-Smith, L., \& Smith,
	R. C. 2002, AJ 123, 255
\bibitem[]{} Oey, M. S. \& Massey, P., 1995, ApJ, 452, 210
\bibitem[]{} Oey, M. S. \& Smedley, S. A. 1998, AJ, 116, 1263
\bibitem[]{} Oey, M. S., Watson, A. M., Kern, K., \& Walth,
  G. L. 2005, AJ 129, 393
\bibitem[]{} Palou\v s, J., Franco, J., \& Tenorio-Tagle, G. 1990,
  A\&A 227, 175
\bibitem[]{} Parizot, E., Marcowith, A., van der Swaluw, E., Bykov,
  A. M., \& Tatischeff, V. 2004, A\&A 424, 747
\bibitem[]{} Pikel'ner, S. B. 1968, Astrophys. Lett., 2, 97
\bibitem[]{} Rhode, K. L., Salzer, J. J., Westpfahl, D. J., \& Radice,
  L. A. 1999, AJ 118, 323
\bibitem[]{} Saken, J. M., Shull, J. M., Garmany, C. D., Nichols-Bohlin, J.,
	\& Fesen, R. A. 1992, ApJ, 397, 537
\bibitem[]{} Silich, S. A., Franco, J., Palou\v s, J., \&
  Tenorio-Tagle, G. 1996, ApJ, 468, 722
\bibitem[]{} Silich, S. A. \& Oey, S. 2002, in {\sl Extragalactic Star
	Clusters,} eds. D. Geisler, E. K. Grebel, \& D. Minniti, 
	(San Francisco:  ASP), 459
\bibitem[]{} Silich, S. A., Tenorio-Tagle, G., Terlevich, R., Terlevich, E.,
	\& Netzer, H. 2001, MNRAS, 324, 191
\bibitem[]{} Staveley-Smith, L., Sault, R. J., Hatzidimitriou, D., Kesteven,
	M. J., \& McConnell, D. 1997, MNRAS 289, 225
\bibitem[]{} Stewart, S. G., et al. 2000, ApJ 529, 201
\bibitem[]{} Strickland, D. K. \& Stevens, I. R. 2000, MNRAS, 314, 511
\bibitem[]{} Tenorio-Tagle, G., R\'o\.zyczka, M., Franco, J., \&
  Bodenheimer, P. 1991, MNRAS 251, 318
\bibitem[]{} Thilker, D. A., Braun, R., \& Walterbos, R. A. M. 1998,
  A\&A 332, 429
\bibitem[]{} Thronson, H. A., Lada, C. J., \& Hewagama, T. 1985, ApJ 297, 662
\bibitem[]{} Tomisaka, K. 1990, ApJ 361, L5
\bibitem[]{} Walborn, N. R. \& Parker, J. W. 1992, ApJ 399, L87
\bibitem[]{} Wang, Q. \& Helfand, D. J. 1991, ApJ, 373, 497
\bibitem[]{} Williams, J. P., Blitz, L., \& Stark, A. A. 1995, ApJ
  451, 252

\end{thebibliography}
\end{document}